\documentclass[12pt]{article}
\def\lesssim{\ \hbox{\raise 2pt \hbox{$<$} \kern -13pt
                     \lower 3pt \hbox{$\sim$}}\ }
\def\greatersim{\ \hbox{\raise 2pt \hbox{$>$} \kern -13pt
                     \lower 3pt \hbox{$\sim$}}\ }
\def \pom {{\scriptscriptstyle \kern -0.1em I \kern -0.25em P}}
\begin{document}

\begin{center}
{\Large 
Diffractive jet production in lepton-hadron collisions}

\vskip 50pt
F. Hautmann\\
\vskip 10pt
Physics Department, University of Oregon\\ 
Eugene OR 97403, USA\\ 
\vskip 10pt
and\\  
\vskip 10pt
Institut f{\" u}r Theoretische Physik, Universit{\" a}t Regensburg\\
 D-93040 Regensburg, Germany\\
\vskip 50pt
\noindent{\bf Abstract}
\vskip 20pt
\end{center}

\noindent{
 We study  jet 
 final states in diffractive deep-inelastic scattering.  
 We   present     the  QCD   factorization formula  
 in terms of 
  diffractive parton distributions and discuss  
  its implementation  in  
  NLO  Monte Carlo generators. We  
   compute NLO predictions for the 
 diffractive  jet cross section.  We use this 
  calculational framework to    discuss  
 theoretical models of the long-distance behavior.         
} 

\newpage

\noindent  {\Large \bf 1. Introduction} 

\vskip .7 true cm

 Among the most intriguing results  that have come from 
the HERA  lepton-hadron  collider are 
observations of diffraction hard  scattering~\cite{abra}.  
In these processes, while 
 the     short-distance behavior   is the one well known from   
  ordinary deep inelastic scattering (DIS),    
 the  long-distance  matrix elements   are much more complex 
  than  the usual parton 
 distributions~\cite{bere,hksnpb} and fully incorporate  the  
 dynamics of hadron diffraction. 

These  diffractive  matrix elements, or 
diffractive parton distributions, 
 have so far 
been  considered  mostly in the context of 
 structure functions. 
One of their key properties, however,   concerns 
the size of the gluon distribution, which,  
in sharp contrast with the case of 
 ordinary DIS,  
  dominates    
 the  quark distributions  
 by one order of magnitude    
even at  low mass scales and large momentum fractions.  
This   enhances 
the importance of processes, such as   the  production of jets,   
 that unlike structure 
functions  couple directly to  gluons.

The purpose of this paper is to start to   
  analyze systematically the jet structure of 
 diffractive final states  
 in terms of  diffractive parton distributions and 
 next-to-leading-order (NLO)  short-distance cross sections.  
A generic  jet observable $\sigma$
  will be   represented schematically as   
\begin{equation} 
\label{obs}
\sigma [W] = \int    f^{(D)} \otimes H \otimes W \hspace*{0.3 cm} , 
\end{equation}
where $f^{(D)}$ is the diffractive parton distribution, 
$H$ is the  hard scattering function, calculable as a power 
series expansion in $\alpha_s$, and $W$  is the  measurement function,  
 specifying  the definition of the observable in terms   
of the final-state kinematic variables. 
QCD analyses of  
  diffractive jet production have 
so far been limited 
to the     leading logarithms~\cite{bart,h1diffjet,zeusdiffjet,ingel}.  
In this paper,  
we present the  QCD  factorization formula  
that allows nonleading corrections 
to be systematically included and we 
carry out the calculation at the NLO.  
We    find that 
 contributions beyond the leading logs  are  important 
in the HERA kinematic region.  

We  argue  that,  
once  perturbative corrections are consistently taken into account, 
jet production   can be used to test  theoretical models of  the 
long-distance behavior  of   
diffraction scattering.  
Motivated by a 
previous 
comparison~\cite{hscoltrans} of theory with  
diffractive $F_2$ data~\cite{h1f2,zeusf2}, 
we consider a simple physical picture~\cite{hksnpb,hscoltrans} in which  
the diffractive 
gluon distribution  results predominantly from  
   color-octet dipoles  penetrating 
  the target at small transverse separations   of order   
  $1 / \kappa$, where 
  $1 / \kappa$ represents 
  a color transparency length. This  picture gives testable predictions, 
   because the smallness of $1 / \kappa$ justifies a perturbative 
 calculation of the $\beta$ dependence of the distributions. 
We compute  predictions 
for jet cross sections in this scenario.

The paper is organized as follows. 
We start in Sec.~2 by presenting the factorization formula  for 
diffractive jet production  and  
its implementation    in 
NLO Monte Carlo generators. 
 This discussion provides   a ``$t$-channel''  picture of  
the process,  based on the separation of 
 high and low transverse momenta at  scale 
$\mu$, and renormalization-group evolution in $\mu$.  
In  Sec.~3 we  examine 
  the diffractive parton distributions.  
  To 
 gain insight into  the form of these  
distributions,  it  is  useful to exploit 
the intuition that comes from  an 
``$s$-channel''  picture    of  the process,  based on the 
 target  rest frame  and the 
light-cone 
evolution of the parton system created by the operator that 
defines the distribution.  
In Sec.~4  we use the results of Secs.~2 and 3  to 
compute the diffractive jet cross section,  and 
 present numerical results  at the NLO.  
Due to our lack of familiarity with the kinematic cuts employed in 
the experiments,  
we leave to the experimental 
groups the comparison   of  these results  with the  
  data. 
We  summarize and give concluding remarks  in Sec.~5.

\vskip 1.5 true cm 

\noindent  {\Large \bf 2. NLO  event generators and hard  diffraction} 

\vskip .7 true cm

In this section we discuss  the diffractive  factorization 
formula. Using properties of this formula, we 
adapt standard NLO Monte Carlo generators for DIS  in order to 
perform NLO calculations of  diffractive hard processes. 

Consider the diffractive leptoproduction of 
 jets, $e  + A \to e^\prime + A^\prime + J + X$,   by 
virtual photon exchange. 
We  parameterize 
 the momenta of the incoming 
hadron  and  virtual photon 
 in a collinear reference frame as 
\begin{equation}
\label{collframe}
p_A = \left(  p_A^+  , {m^2_A \over 
{2   p_A^+} }  , {\bf 0} \right)
 \hspace*{0.3 cm} ,  
 \hspace*{0.6 cm}
  p_\gamma = \left( - x  p_A^+  , {Q^2 \over 
{2  x  p_A^+} }  , {\bf 0} \right) 
 \hspace*{0.3 cm} ,    
\end{equation} 
where $Q^2$  is the photon virtuality,   $x$  is the Bjorken variable,  
 and we  have used  
lightcone  components  $p^\pm = (p^0 \pm p^3 ) / \sqrt{2}$. 
The Breit frame, in 
which we will  define the jets, 
 is identified by    $p_A^+ = Q / (x \sqrt{2})$. 
We  set    
$y= Q^2/(x S)$, with $S$ the lepton-hadron 
center-of-mass energy squared.    
Hadron $A^\prime$ is characterized 
by the 
  fractional loss 
of longitudinal momentum  $x_\pom = 1 -  p_{A^\prime}^+  / p_A^+$    
and by the invariant momentum transfer $t = (p_A - p_{A^\prime})^2$.

To be definite,  we suppose   fixing   $Q^2$ and $y$ in the 
lepton subprocess,  
$x_\pom$ in the 
diffractive subprocess, 
and the transverse energy $E_T$ of the jet, whose precise definition 
is specified below (see Sec.~4). 
We integrate over $t$ up to a maximum value 
$t_{\rm{max}}$, which we 
 take to be  of the order of a  $ {\mbox {GeV}}$.  
We   consider 
  the fourfold-differential jet cross section 
 $d\sigma$$/$$[dE_T 
 dQ^2  dy dx_\pom ]$.  According to 
 the hard-scattering 
 factorization theorem~\cite{bere,hksnpb,graufrac,diffcontour}, 
   this cross section 
is given, 
 up to corrections suppressed by  powers of the 
hard-scattering   scale,  by 
\begin{eqnarray} 
\label{factsig} 
&& {{d \sigma 
} \over { dE_T  dQ^2  dy dx_\pom  }} 
( E_T , Q^2 , S, y, x_\pom )
 =  \sum_{a= g , q , {\bar q}} 
\int_\beta^1 { {d \beta^\prime} \over \beta^\prime}  
\nonumber\\ 
&& \times 
{{d f^{(D)}_{a}} \over {  dx_\pom }} 
(\beta^\prime  , x_\pom , \mu^2) \ 
{{d {\hat \sigma}_{a}} \over {d E_T  dQ^2  dy   }} 
(\beta / \beta^\prime , E_T, Q^2 , y , \mu^2, 
\alpha_s(\mu_R^2) ) \hspace*{0.3 cm} ,    
\end{eqnarray} 
where  $\beta$ is given by 
\begin{equation}
\label{betakin}
\beta = {{Q^2} \over { x_\pom y S}}  
 \hspace*{0.3 cm} ,      
\end{equation}
${d f^{(D)}_{a}} /   dx_\pom $ is  the diffractive parton 
distribution integrated over $t$,  and 
${d {\hat \sigma}_{a }} /$  $[d E_T  dQ^2  dy   ]$ 
is the partonic jet cross section,  incorporating  the 
hard-scattering 
function $H$ of Eq.~(\ref{obs})  
(through perturbatively calculable  matrix 
elements) and the measurement function 
$W$ (through the algorithm that defines  the jet). 
The mass scales  $\mu_R$ and $\mu$ are the 
renormalization and   factorization scales.

Eq.~(\ref{factsig})  expresses 
 the 
 separation of 
short-distance  
and long-distance 
contributions to the diffractive jet cross section. It  contains  
 two crucial differences compared to the 
analogous factorization formula for  inclusive jet  
production: a) the distributions  $f^{(D)}$, which 
embody all of the dynamical effects of hadron diffraction, and 
b)  the diffractive kinematics, which enters, through  
 the  momentum loss $x_\pom$,  not only in 
$f^{(D)}$  but also in $\beta$,  Eq.~(\ref{betakin}). 
The    definition 
   of   $f^{(D)}$  
   can be found in \cite{bere,hksnpb} 
  and  is  briefly recalled in Sec.~3. 

 The important  point for practical applications  is that the 
 dependence on the hard  scale,   
  $Q$ or $E_T$, factors out 
 of $f^{(D)}$ in Eq.~(\ref{factsig}): it is entirely contained in  
 ${\hat \sigma}_{a }$, as in the inclusive case. 
 Since the left hand side of Eq.~(\ref{factsig}) is independent of 
the factorization scale $\mu$, this also implies that, as in 
the inclusive case, the  evolution of  $f^{(D)}$ with $\mu$ 
is given by  renormalization-group  equations. 
Because we do not  integrate over large $t$ (of the order of the hard 
scale or higher), 
the form of these equations is  simply DGLAP, without the 
inhomogeneous term 
from  extra large-$t$ subtractions~\cite{bere}. 

 We can thus calculate 
 the differential cross section (\ref{factsig}) to the 
 next-to-leading  order using  the NLO results that are 
  available for  the partonic  jet  cross sections ${\hat \sigma}_{a}$, 
  and  NLO evolution of the $f^{(D)}$.     
The  cross sections 
${\hat \sigma}_{a}$  
start at  order  $\alpha_s$. 
The order-$\alpha_s^2$ contributions  
were computed   
and encoded in Monte Carlo programs 
by 
three groups~\cite{graudisa,catsey,mirzepp}. Two of these programs,   
 DISASTER++~\cite{graudisa} and  DISENT~\cite{catsey}, 
have been found~\cite{compnlo} to be 
   in good 
numerical agreement. More 
recently  
two other  independent  programs~\cite{poett,nagy} have appeared.  
The authors of  the program NLOJET++~\cite{nagy} have also performed
 a comparison of different codes 
and found agreement of their results with those 
of \cite{graudisa,catsey}. 
Any one of these  Monte Carlo programs can be 
used to evaluate  the short 
distance part of Eq.~(\ref{factsig}) to NLO~\footnote{    
Note that  
 a  numerical discrepancy 
 between  DISASTER++ and  DISENT 
   has been observed~\cite{compshape} 
 in the case of certain event shape 
 distributions,  e.g.    
 the jet broadening, and that  
 DISASTER++ has been identified~\cite{dasgusal} as the program  
 giving the correct  result. 
 As far as we can judge, this 
 discrepancy does not affect the cross sections and the range in 
 transverse energies   considered in this paper.}. 

 The numerical results that we present in this paper 
 are  obtained using  DISENT.   
This program generates DIS events 
according to a structure reminiscent of 
 parton shower algorithms~\cite{seyweb}: 
first,  
it generates 
a parton-model event,   with one parton in the 
final state;  then, starting from this and 
using the dipole subtraction method~\cite{catsey},  
it generates 
events with two  partons in the 
final state; then, in a similar way 
it generates 
 events with three   final partons.  
 Contributions to the jet cross section come from events with  
 at least two partons in the final state. 
 For applications to diffractive production,  we  generate 
  two-parton and three-parton events so that $Q^2$ and $\beta$ of the 
  starting event are preserved.      
 Note that if  we use variables 
 $Q^2$,  $y$ and $\beta$ for generating 
  the DIS phase space (any two of which can be 
  fixed),  the  dependence  on the diffractive 
  momentum loss  $x_\pom$ 
 decouples from the convolutions 
 implied by  factorization and evolution. 
We will use this method for the calculations of Sec.~4.

In the next section  we  turn to   the 
long-distance part of the process. 

\vskip 0.5 true cm 

\noindent  {\Large \bf   3. Diffractive parton distributions  } 

\vskip .2 true cm

The diffractive parton distributions 
describe the nonperturbative  dynamics 
of the hadronic state, and    
  are  defined  in  \cite{bere,hksnpb} in terms of  
matrix elements of nonlocal quark and gluon 
field operators. 
For gluons one has 
\begin{eqnarray}
\label{opg}
&& {{d\, f_{g}^{(D)} } \over
{dx_\pom}} (\beta , x_\pom ,  \mu^2 ) =  
{1 \over (4\pi)^3 
\beta x_\pom p_{\!A}^+}\sum_{X} \int_0^{t_{\rm{max}}} dt \int d y^-
e^{i\beta x_\pom p_{\!A}^+ y^-} 
\nonumber\\
&&   \times 
\langle A |\widetilde G_a(0)^{+j}
| A^\prime, X \rangle 
\langle A^\prime, X|
\widetilde G_a(0,y^-,{\bf 0})^{+j}| A \rangle 
\hspace*{0.2 cm} ,     
\end{eqnarray}   
with 
\begin{equation}
\label{Gop}
\widetilde G_a(y)^{+j}
=
E(y)_{ab}
G_b(y)^{+j} \hspace*{0.2 cm} ,  \hspace*{0.2 cm} 
E(y) = 
{\cal P}
\exp\left(
- i g \int_{y^-}^\infty d x^-\, A_c^+(y^+,x^-,{\bf y})\, t_c
\right)   \hspace*{0.2 cm} ,  
\end{equation}
where $G$ is the gluon field strength, $A$ is the 
vector potential, $t$ are the generators of the adjoint representation of 
$SU(3)$, and $E$ is a color-octet eikonal line operator, with 
${\cal P}$ denoting the  path ordering of the exponential.  
   An analogous definition  applies for  quarks. 
Unlike the  ordinary (inclusive) parton distributions,
the  distributions  $f^{(D)}$ 
 represent   interactions that occur both  long before and 
 long after the hard   scatter.

 The   factorization   of the previous section   
implies that the $f^{(D)}$'s  can be measured from 
diffractive DIS data and used in 
 Eq.~(\ref{factsig}) to predict  jet cross sections. 
  This   provides   an    approach  to diffractive 
 jet production that is fully consistent,   although 
 agnostic about the form of the 
matrix elements (\ref{opg}),  
 in the spirit of the parton model.  
  Calculations based on this approach   are 
   carried out    to leading order in  
   \cite{h1diffjet,zeusdiffjet} and  
   to NLO in  this paper.  
    From this point of view,  experimental 
  tests of   Eq.~(\ref{factsig})  to NLO  are an important goal 
of diffractive  jet studies.

A  complementary approach  consists of 
modeling  the dynamics that determines the form  
of the matrix elements (\ref{opg}),  
and    
  using  
 Eq.~(\ref{factsig}) 
to test  these models.  It is this point of view 
  that we  now wish  to take. 
To this end,  we build on the observation~\cite{hksnpb,hkslett}  that 
 the main qualitative aspects of the 
 data~\cite{h1f2,zeusf2} for 
 the diffractive $F_2$  structure function 
 --- steep  rise with decreasing $x_\pom$,    
flat spectrum  in $\beta$,   
positive slope in  $Q^2$ up to 
$\beta \approx 1/2$ --- 
are all   consistent with the hypothesis  that the 
diffractive gluon distribution be dominated by transverse 
momenta of the order 
of a  (semi)hard  scale $M_{\rm{SH}}$, 
\begin{equation}
\label{shscale}
 M_{\rm{SH}}   \sim {\cal O} ( 1 \ {\mbox{GeV}} ) 
\hspace*{0.1 cm}  .        
\end{equation}   
This scale has nothing  to do with the hard-scattering scale, $Q$ or 
   $E_T$, which has been factored out and does not appear in $f^{(D)}$. 
   Rather, 
   it represents an intermediate scale between the scale of hard physics 
   and $\Lambda_{\rm{QCD}}$,  whose origin 
    is nonperturbative 
   and associated with the hadron's soft color field.

   The physical meaning of the scale $M_{\rm{SH}}$ 
   was explained in \cite{hscoltrans} in the 
   reference  frame in which the target is at rest.  
   In this frame, the scale  $M_{\rm{SH}}$  is seen to arise 
   as a color-transparency scale, i.e., as the inverse of 
  the maximum  size $1 / \kappa$ for which  the   
  color-octet dipole system created by the operator (\ref{opg})
  can  go 
  through the hadron without breaking it up.   
The  analysis  \cite{hscoltrans} 
 finds that 
 the 
 transparency lengths are smaller for color-octet dipoles than for 
color-triplet dipoles.  
The hypothesis  of semihard dominance  
may therefore be better verified  for observables (such as jets)  
that couple  directly to  the gluon 
distribution  than for structure functions.

Note that the  rest frame 
  approach    gives  an 
``$s$-channel view'' of the process, 
characteristic of   color-dipole models.  
 Note however an  essential 
 difference between the treatment   above and  
 standard color-dipole  models (see, e.g., Ref.~\cite{bart} 
 for a color-dipole study of 
diffractive jet production): in 
 the present paper    the 
 $s$-channel picture  is applied to the parton system created by 
 the operator in $f^{(D)}$ rather than to the system 
 into which the virtual photon  dissociates.   That is, it is applied 
 {\em after} factorizing   the hard-scattering subgraph.  
This 
difference is relevant, because the factorization 
provides a framework in which 
 next-to-leading radiative corrections 
and evolution can be  readily implemented.  
In contrast, such  effects are difficult  to include  in 
color-dipole models, and 
evolution is typically taken into account   
at most   to the leading-log level: see, e.g., \cite{golekowa} 
for a recent study.

The    semihard dominance  picture 
leads to testable predictions, because  
   the $\beta$ dependence of the diffractive parton distributions 
 then becomes  calculable by a 
perturbation expansion.  
This calculation  was first done in \cite{hkslett}, using 
light-cone perturbation theory. 
We observe that the semihard-scale  scenario 
is common to several different models, including 
the large-nucleus model~\cite{buchmue}  and  the 
saturation model~\cite{golec}.  
In the  following  we 
use the  
 distributions of \cite{hscoltrans}, which incorporate the 
    calculation~\cite{hkslett}.     
At a starting  mass scale    $ \mu=\mu_0$ (with    $ \mu_0$ of order 
$M_{\rm{SH}}$), these distributions   have the form 
\begin{equation}
\label{gform}
{{d f^{(D)}_{g}} \over {  dx_\pom }} 
(\beta  , x_\pom , \mu^2_0) 
   = A_g  \ {1 \over 
x_\pom^{2 \alpha -1} } \   
 \varphi_g (\beta) 
\hspace*{0.1 cm}  , 
\hspace*{0.3 cm}
{{d f^{(D)}_{q}} \over {  dx_\pom }} 
(\beta  , x_\pom , \mu^2_0)  = A_q  \ {1 \over 
x_\pom^{2 \alpha -1} } \ 
 \varphi_q (\beta) 
\hspace*{0.1 cm}  ,         
\end{equation} 
where the   $\varphi$'s  are  
the 
perturbatively-computed  functions, while 
$A_g$,  
$A_q$, $\alpha$ have to be 
determined from 
the data. 
The overall normalizations 
 are proportional to the square 
of the  
color-transparency scales,   
$A_g \propto  \kappa_g^2  r_A^2 $ and   
$A_q \propto  \kappa_q^2  r_A^2 $~\cite{hscoltrans}, 
with $r_A$ the  hadron radius. 

In Eq.~(\ref{gform}) 
we have assumed, as is commonly done, 
 a simple 
factorizing 
   form $x_\pom^{1-2 \alpha}$ 
   for the $x_\pom$ dependence.   
This is an ansatz, not a  theory result. It is 
in fact  possible that the   gluon and the quark 
distributions have distinct  $x_\pom$ behaviors  at low 
$\mu$~\cite{gluepsi}. 
For illustrative purposes  in 
 the calculations that follow  we limit ourselves to the 
simple factorizing form.

Observe that the larger $\kappa$, the larger the parton 
distribution:    large 
  $f^{(D)}_g$  is related to the 
small size of the color-octet dipoles 
interacting diffractively~\cite{hscoltrans}. 
The calculation of the functions $\varphi_g$ and  $\varphi_q$  shows that 
 the ratio of 
the gluon and quark distributions   
is proportional to a large color factor,  
$C_A^2 (N_c^2 -1 )/ (C_F^2 N_c) = 
27/2$~\cite{hkslett}. In  contrast with ordinary DIS, 
 these functions 
 do {\em not} have a fast fall-off as $\beta \to 1$, 
resulting in  the gluon distribution being large even 
 at large $\beta$~\cite{hkslett}.

As we have seen in Sec.~2, 
the evolution of the distributions (\ref{gform})  with the scale 
$\mu^2$ 
is given, up to power-like corrections,  
by the DGLAP equations, and 
is  to be computed  to the NLO, 
consistently with the  accuracy of the partonic 
cross section.  We do this using the moment-space 
evolution program of \cite{ehwevol}. Numerical tables for the 
resulting distributions in the ${\overline{\mbox{MS}}}$ scheme 
are available from 
http://zebu.uoregon.edu/${\tilde{~}}$parton/diffpartons. 

The use of the DGLAP  
 evolution equations is consistent with the 
leading power accuracy of Eq.~(\ref{factsig}).  
However,  as emphasized 
earlier,    the diffractive gluon distribution $f^{(D)}_g$ 
is very large.  Then  
corrections to factorization and evolution 
of relative order $f^{(D)}_g / ( r_A Q)^n$,  
corresponding to multi-parton correlations,  
although power-suppressed 
should likely be 
significant.  
 Diffractive final states 
 are  therefore   an especially important   case in which  to 
 look  for  signals of  behaviors beyond the leading power.

\vskip 1.5 true cm 

\noindent  {\Large \bf   4. Jet cross sections } 

\vskip .7 true cm

In this section we   calculate the  cross section 
for  diffractive leptoproduction  of jets by  evaluating  
Eq.~(\ref{factsig}) at the NLO.  
For the short-distance part of the process,  we use 
the Monte Carlo program DISENT as discussed in Sec.~2. 
For the long-distance part, we  use the 
diffractive parton distributions described in Sec.~3. 

We must specify the prescription  for 
 converting the final-state partons into jets. 
Beyond the leading logarithms,   
 this  
  becomes an essential ingredient 
 of Eq.~(\ref{factsig}): it  is part of the 
 function $W$ in  Eq.~(\ref{obs}). 
Standard, infrared-safe definitions are available 
both for cone jet algorithms and  for $k_\perp$-clustering jet 
algorithms:  
see, e.g.,  \cite{run2jet} for a recent, comprehensive  discussion. 
In the case of  inclusive jet 
measurements~\cite{h1incljet} in DIS, 
 the H1 Collaboration have performed a comparison of 
 different $k_\perp$  algorithms and found  
the    algorithm of \cite{ellsop,catalgo}
  to  have the smallest hadronization corrections. 
  Since    
  we do not expect any dramatic difference    
  from  the diffractive case in this respect, 
we follow this observation and 
present results for jets   defined using this algorithm. 
As in \cite{h1incljet}, 
we adapt the algorithm, 
originally defined for hadron-hadron collisions, 
 to the lepton-hadron case,  working in the  
Breit  frame.  
We use the Snowmass recombination prescription~\cite{snow} 
to define the 
transverse energy $E_T$, 
 pseudorapidity $\eta$ and  azimuth  
  $\phi$ of the jet in terms of the momenta of its constituent partons. 
  In what  follows   these  kinematic variables 
are understood to be  in the Breit frame.

\begin{figure}[htb]
\vspace{95mm}
\includegraphics{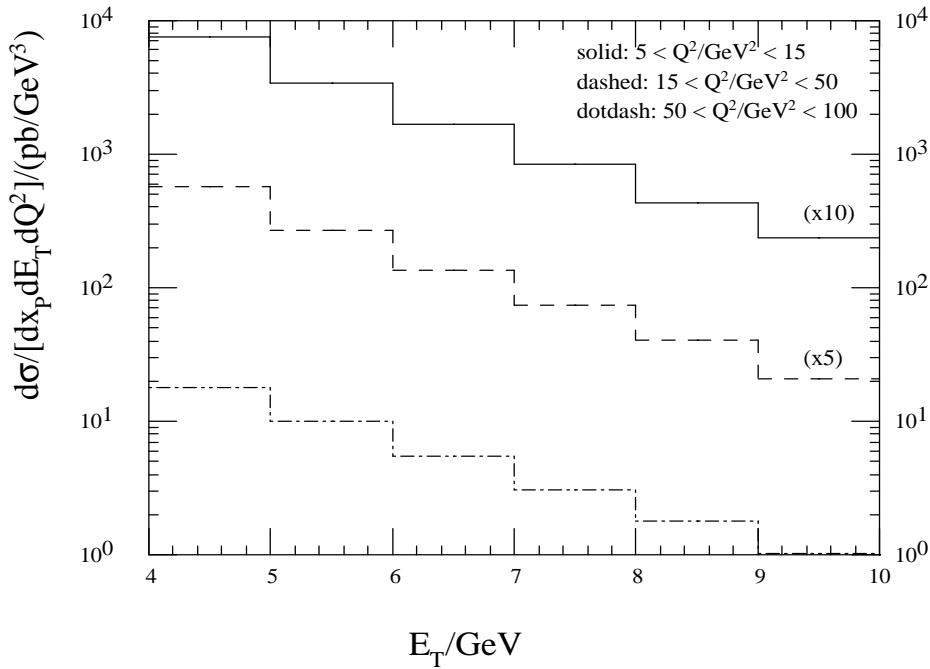}
\caption{The diffractive one-jet cross section at NLO 
as a function of the jet transverse energy $E_T$ for  different 
$Q^2$ bins and  $x_\pom=0.05$ 
($\protect \sqrt S = 300$~GeV). The energy fraction $y$ is 
integrated over the interval $0.1 < y < 0.7$.  
Jets are defined by $k_\perp$-clustering, as specified in the text.  }
\label{firstfiget}
\end{figure}

\begin{figure}[htb]
\vspace{95mm}
\includegraphics{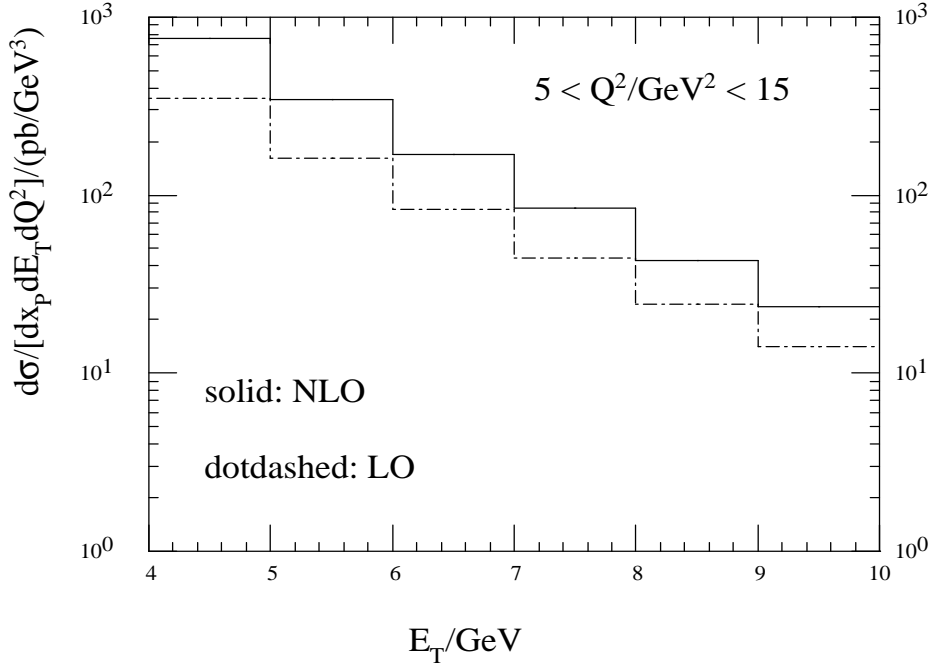}
\caption{Size of the NLO 
correction. The values 
of the kinematic variables 
and range of integration  are as in  Fig.~1 (upper curve).}
\label{secondfig}
\end{figure}

\begin{figure}[htb]
\vspace{95mm}
\includegraphics{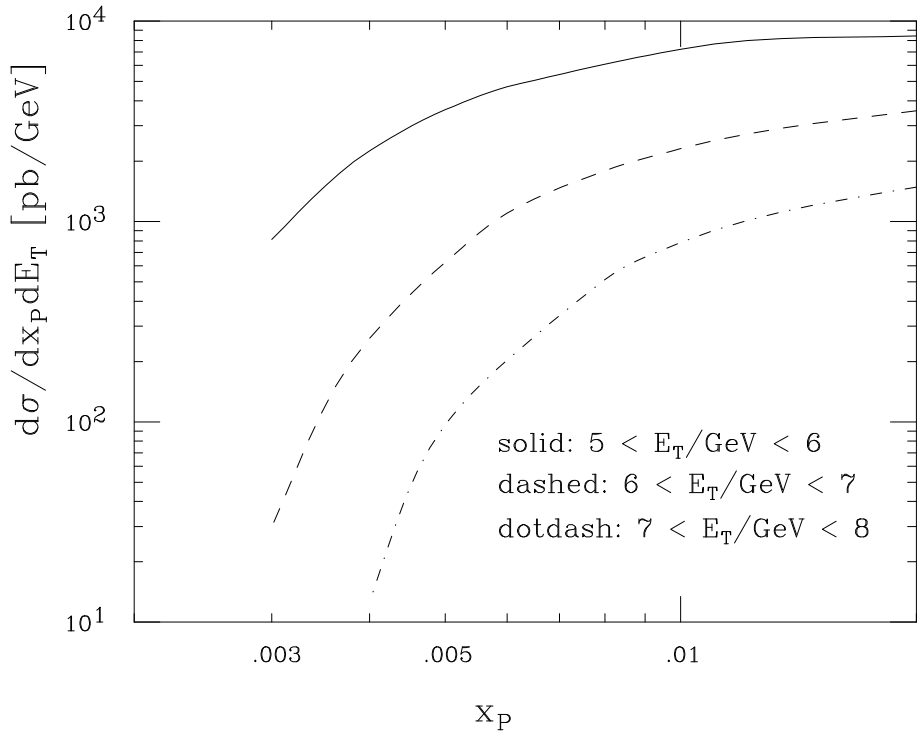}
\caption{
The diffractive one-jet cross section at NLO 
as a function of the hadron's momentum loss 
$x_\pom$ for different  jet transverse energies $E_T$.    
$\protect \sqrt S = 300$~GeV;  
$4$   GeV$^2$ $< Q^2 < 70$ GeV$^2$;  
 $0.1 < y < 0.7$.  
 Jets are defined by $k_\perp$-clustering, as specified in the text.  }
\label{firstfigxp}
\end{figure}

In the calculations that we present below 
we integrate the momentum transfer $t$ from $t =0$  to a maximum value 
$t_{\rm{max}} = 1 \ {\mbox {GeV}}$.  
 We set the  
 factorization and renormalization 
  scales  to  $\mu^2 = \mu^2_R = Q^2$.  
We set the jet resolution parameter $R$~\cite{ellsop} to the value $R=1$.

In Fig.~1 we integrate the cross section (\ref{factsig}) over 
$y$, with $0.1 < y < 0.7$, and plot   
 the  
triple-differential cross section 
$ d \sigma / [ dE_T 
 dQ^2  dx_\pom ]$ versus $E_T$ at a fixed value of 
  $x_\pom $  for three different bins in $Q^2$. 
  The  values chosen for  the kinematic variables 
 are in the range accessable at HERA. 

In order to isolate the quantitative effect of the 
next-to-leading   correction to the hard scattering, 
 in Fig.~2 we 
  plot the same result 
  as in Fig.~1 for the lowest $Q^2$ bin 
  along with  the  result obtained by including only the 
  leading-order contribution to the hard scattering.  The 
  diffractive parton   distributions  and the value of the 
  running coupling are the same in the two curves of Fig.~2. 
The effect is of the order of a factor of 2. 

Note that 
from the point of view of the $s$-channel picture of the process 
(see,  e.g., 
Ref.~\cite{bart} and discussion in Sec.~3), the LO curve  corresponds to 
the sum of the   contributions 
 $\gamma^* \to q \bar q$ and  $\gamma^* \to q \bar q g$,  
with the latter being by far the  dominant one, since 
$f^{(D)}_{g} \gg f^{(D)}_{q}$. The NLO curve 
corresponds to   the first radiative correction to these two contributions.  
  Fig.~2  indicates  that  
 the  radiative  corrections (besides evolution)  to  
 photon dissociation into  $q \bar q g$ 
 are    important 
 in the HERA region.

\begin{figure}[htb]
\vspace{100mm}
\includegraphics{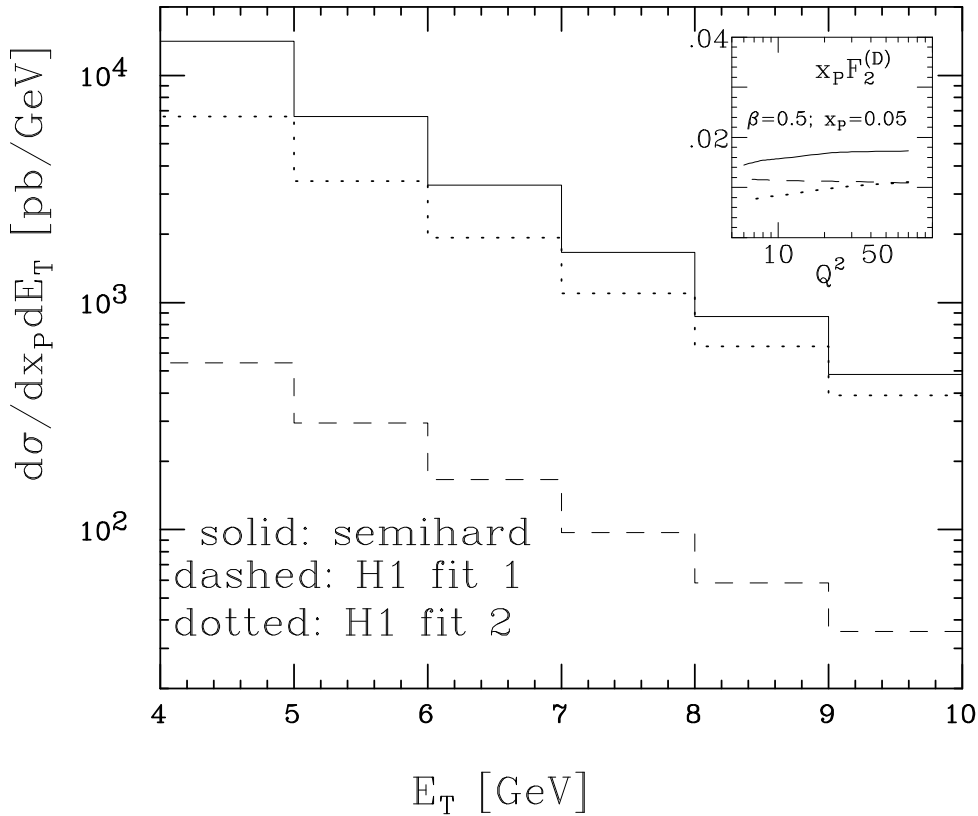}
\caption{Predictions for the jet cross section 
corresponding to different  diffractive parton distributions. 
The distributions in the semihard dominance picture 
(solid curve) are those of \protect \cite{hscoltrans}; 
 the H1 fit distributions (dashed and dotted curves) are from 
 Ref.~\protect\cite{h1f2}. 
  The range of integration in $Q^2$ and $y$ is 
 as in Fig.~3;  $x_\pom = 0.05$.  
 Inset:  
  results  for  the diffractive structure function $F_2$ 
  from the same sets of diffractive parton distributions.
 }
\label{fourthfig}
\end{figure} 

The size of the correction raises the question of 
how 
  to improve the reliability of the perturbation 
series.  
Because the short-distance matrix elements are weighted 
by parton distributions that behave very differently than  the ordinary 
  distributions, 
  different  physical effects  at short distances 
  will  dominate  the correction 
compared to the inclusive case. 
Recall from Sec.~3 that   
$f^{(D)}_g$ stays large up to  large  values of $\beta$.  
Then the physical cross section is likely to be sensitive  
to the  behavior   of the  partonic 
cross section for  small longitudinal momentum fraction and 
to the   tail at  finite  parton transverse momenta.  
We leave the analysis of this issue to future investigation.

 In  Fig.~3  
 we  show  the $x_\pom$ dependence  of the cross section 
  for  different  $E_T$ bins. 
  We integrate 
  the triple-differential cross section defined above 
   over $Q^2$, with $4$   GeV$^2$ $< Q^2 < 70$ GeV$^2$, 
  and plot $ d \sigma / [ dE_T 
   dx_\pom ]$. 
 Although the  distributions $f^{(D)}$  
 increase with decreasing $x_\pom$ --- 
 see  Eq.~(\ref{gform}), with 
 the measured value 
 $\alpha \simeq 1.15$~\cite{h1f2,zeusf2} ---,  
 the 
  jet cross section  decreases 
  as a result of the reduction in 
  the longitudinal phase space ---  
see   
  Eqs.~(\ref{factsig}) and (\ref{betakin}).  
This behavior is due to    the 
non-pointlike coupling  of jets  to the electromagnetic current, and  
 characterizes the measurement  of jets 
with respect  to  the measurement 
 of the 
structure function $F_2$.  

Having  examined the NLO  perturbative corrections, 
let us now look at the impact of long-distance effects on the 
jet cross section.  To illustrate this,  
we make use  of  two fits performed by the  H1 Collaboration to 
structure function data~\cite{h1f2}. 
We compare 
 predictions  based on  the diffractive parton distributions  in 
 the   semihard dominance picture 
    with predictions  
    based on  the diffractive parton distributions 
  extracted from   the  H1  fits.   
  The fit distributions, 
  while  compatible with 
  diffractive $F_2$  data,  
  are 
     representative of   different 
 physical  pictures of the long-distance process. 

 Results    are given  in Fig.~4.  
 We plot the doubly-differential jet cross section 
  defined  above 
  as a function of the jet transverse energy $E_T$.  
The solid lines in Fig.~4 
 correspond to  the same distributions~\cite{hscoltrans}  
as in the previous figures;  the dashed and  dotted lines  
correspond to the fit-1 and fit-2  distributions of H1~\cite{h1f2}.   
(We recall from \cite{h1f2} that 
fit-1 distributions are quark-dominated, while  fit-2 
distributions are gluon-dominated.) The inset in the upper 
right corner of Fig.~4   shows  the 
corresponding NLO results for the diffractive structure 
function $F_2$,  obtained from the same three sets of distributions. 
 The  comparison gives a quantitative illustration of  
  how much more sensitive the jet cross section is to  
 long-distance effects than $F_2$.  
 We may also remark, from the results of Fig.~2 and Fig.~4, that 
 while   perturbative corrections to  diffractive jet rates 
 at HERA are large, on the order of a factor of 2, the effects from 
different  scenarios for the 
parton distributions   (all compatible with 
$F_2$  data)  are even larger,  on the  order of a factor of 10.  
We should likely learn a great deal about the 
 long-distance physics  of  hard  diffraction
from  the study of jet final states.

\vskip 1.5 true cm 

\noindent  {\Large \bf   5. Summary } 

\vskip .7 true cm

In this paper we have presented the factorization formula that relates 
the diffractive jet-production cross section 
in DIS 
to the diffractive parton 
distributions. Previous studies of diffractive jet 
leptoproduction have been 
based on 
approaches that do not go beyond 
the leading logarithm approximation. The factorization formula 
provides a systematic framework that allows arbitrarily nonleading 
corrections to be included. We have evaluated this formula explicitly 
to the next-to-leading order. 

Using factorization,  we have discussed how to adapt 
standard NLO event generators  in order to perform 
NLO calculations for diffractive jet physics.  
We have used this method to compute one-jet cross sections. 
The method is general and can be applied to a 
variety of final-state observables.

This  improved calculational  framework  
 can be used to study the diffractive gluon 
distribution. 
Building on previous work,     
and motivated by indications from diffractive 
$F_2$ data,  we have discussed  
 scenarios for long-distance 
 physics  in which  the  diffractive gluon 
distribution  
is  dominated by 
color-transparency lengths 
 of the order of the inverse of a semihard scale,  
$M_{\rm{SH}} \sim 1$ GeV.  
In this case the $\beta$ dependence 
becomes calculable by   perturbation  methods, and leads to 
testable predictions for the jet cross sections.

We recall that the factorization formula is valid  up to 
corrections suppressed by powers of the hard scattering scale. These 
corrections  
correspond to multi-parton exchanges and  
contributions nonlinear in the 
parton distributions.  
Since   the diffractive  
gluon distribution 
is very large,  this may  overcome 
the power suppression. We  underline  the importance of  
searching  for deviations from 
the leading power particularly in  diffractive final states.

We have observed  that  NLO contributions to 
the diffractive cross sections  are generally large 
in the HERA kinematic region.  
In  the $s$-channel language in which color-dipole models are 
most naturally formulated, this indicates that  
nonleading-log corrections to 
 photon dissociation into  $q \bar q g$ 
states  are important. 
We  also noted the possibility that 
 effects   from finite parton transverse momenta  in the 
 short-distance cross section   
may be exposed by the large-$\beta$ behavior of the 
diffractive  gluon distribution. 
 The investigation of these questions  is left to future work.

\vskip 1.5 true cm 

\noindent  {\Large \bf  Acknowledgments} 

\vskip 0.7 true cm 

I am grateful to  D.~Soper for collaboration on 
diffractive physics and  for his invaluable advice. 
I   acknowledge  discussions with 
V.~Braun, S.~Catani, J.~Collins, Z.~Kunszt and M.~Strikman.  
This work is funded in part by the US Department of Energy.

\end{document}